\documentclass[journal]{IEEEtran}

\usepackage{setspace}
\usepackage{amsmath}
\usepackage{amsfonts}
\usepackage{graphicx}
\usepackage{amsthm}
\usepackage{amssymb}
\usepackage{mathrsfs}
\usepackage{stfloats}
\usepackage{geometry}
\usepackage{color}
\begin{document}
\title{On Performance of Quantized Transceiver in Multiuser Massive MIMO Downlinks}


\author{
%\IEEEauthorblockN{Author 1, Author 2, and Author 3}
%\IEEEauthorblockN{Jindan Xu$^{1,2}$,~\emph{Student Member,~IEEE}, Wei Xu$^{1,2}$,~\emph{Senior Member,~IEEE}, and Fengkui Gong$^{2}$,~\emph{Member,~IEEE}}
\IEEEauthorblockN{Jindan Xu,~\emph{Student Member,~IEEE}, Wei Xu,~\emph{Senior Member,~IEEE}, and Fengkui Gong,~\emph{Member,~IEEE}}
%\IEEEauthorblockA{$^{1}$National Mobile Communications Research Laboratory, Southeast University, Nanjing 210096, China.\\
%$^{2}$State Key Laboratory of ISN, Xidian University, Xi¡¯an, China.}
%\IEEEauthorblockA{Email: \{jdxu, wxu\}@seu.edu.cn, fkgong@xidian.edu.cn}
\vspace{-0.8cm}
\thanks{
Manuscript received April 21, 2017; revised May 27, 2017; accepted June 13, 2017.
This work of was supported in part by the 973 program 2013CB329204, the NSFC under grants 61471114, 61372067, and 61571118, the Open Research Fund of the State Key Lab of ISN under ISN18-03, and the Six Talent Peaks project in Jiangsu Province under GDZB-005.
The editor coordinating the review of this paper and approving it for publication was A.Kammoun.
\textit{(Corresponding author: Wei Xu).}

J. Xu and W. Xu are with the National Mobile Communications Research Laboratory (NCRL), Southeast University, Nanjing, China (jdxu@seu.edu.cn; wxu@seu.edu.cn).
They are also visiting scholars with the State Key Laboratory of Integrated Services Networks, Xidian University, Xi¡¯an, China.

F. Gong is with the State Key Laboratory of Integrated Services Networks, Xidian University, Xi¡¯an, China (fkgong@xidian.edu.cn).

}
}


\maketitle

\begin{abstract}
Low-resolution digital-to-analog converters (DACs) and analog-to-digital converters (ADCs) are considered to reduce cost and power consumption in multiuser massive multiple-input multiple-output (MIMO). Using the Bussgang theorem, we derive the asymptotic downlink achievable rate w.r.t the resolutions of both DACs and ADCs, i.e., $b_{DA}$ and $b_{AD}$, under the assumption of large antenna number, $N$, and fixed user load ratio, $\beta$. We characterize the rate loss caused by finite-bit-resolution converters and reveal that the quantization distortion is ignorable at low signal-to-noise ratio (SNR) even with low-resolution converters at both sides. While for maintaining the same rate loss at high SNR, it is discovered that one-more-bit DAC resolution is needed when more users are scheduled with $\beta$ increased by four times.
More specifically for one-bit rate loss requirement, $b_{DA}$ can be set by $\left\lceil b_{AD}+\frac{1}{2}\log\beta \right\rceil$ given $b_{AD}$. Similar observations on ADCs are also obtained with numerical verifications.
\end{abstract}

\begin{IEEEkeywords}
Massive MIMO, DAC, ADC, quantization.
\vspace{-0.3cm}
\end{IEEEkeywords}

%\vspace{-0.3cm}

\IEEEpeerreviewmaketitle

\section{Introduction}
Massive multiple-input multiple-output (MIMO) has become a candidate technique for the fifth generation (5G) wireless communication system \cite{MIMO1}. Despite its several advantages \cite{MIMO3}, nevertheless, hundreds of antennas significantly increase cost and power consumption partly because each antenna requires a digital-to-analog converter (DAC) unit at transmitter or an analog-to-digital converter (ADC) at receiver. Currently, there exist two kinds of approaches for potential solutions. One is to exploit hybrid precoding with a very limited number of radio-frequency (RF) chains serving all antennas, like in \cite{Hybrid2}. Alternatively, each antenna element is as usual connected with a dedicated RF chain which is, however, served by low-resolution, or even 1-bit, ADC/DACs, e.g., \cite{DAC1}.

%On one hand, the number of radio frequency (RF) chains is reduced and hybrid precoding is proposed \cite{Hybrid1}. By this way, the number of devices required remarkably decreases. On the other hand, low-resolution DAC or ADC is used in each RF chain \cite{ADC1} and the power consumption of each device is reduced. \cite{Hybrid ADC} combined these two kind of methods. There are tradeoffs between system performance and power consumption in all the above schemes.

In particular, \cite{ADC1} showed that 1-bit ADCs achieve satisfactory performance in uplink massive MIMO. Mixed ADCs were considered in \cite{ADC2}, which revealed that the quantization distortion caused by low-resolution ADCs can be compensated logarithmically by more antennas. Recently, some researchers studied quantized precoding with low-resolution DACs for downlinks. \cite{DAC2} considered a special case of 1-bit DACs and showed that the system performance depends on user load ratio. Most existing studies focused only on either low resolution DACs or ADCs at BS side for the sake of tractability. However, finite-bit converters are generally employed at user side.
In a single-antenna system, it is directly known by the Shannon theory that the achievable rate depends only on the lower resolution between transmitting DAC and receiving ADC. For multi-antenna setups, however, the joint design of finite-bit DACs and ADCs, as well as its effect on performance, remains unknown and should intuitively rely on the antenna number, especially for the multiuser massive MIMO case.

%Most existing studies using low-resolution devices focused only on DACs in transmitter or ADCs in receiver. However, optimal bit allocation for both DACs and ADCs is to the benefit of finding a balance between spectral efficiency and power thrift. It is obvious that the achievable data rate depends only on the lower resolution between the transmitting DAC and receiving ADC in a point-to-point scenario where both sides equipped with a single antenna. In other words, the higher resolution device wastes a certain amount of power since it contributes no more benefit to spectral efficiency. However, multiple antennas and users significantly complicate this problem in multiuser massive MIMO.

This letter considers both finite-bit DACs at BS and finite-bit ADCs at user side for downlink massive MIMO. We derive an asymptotic expression for the downlink achievable rate. The rate loss due to transceiver quantization is accordingly characterized. At high signal-to-noise ratio (SNR), the DAC resolution, $b_{DA}$, can be approximately set as $\left\lceil b_{AD}-\frac{1}{2}\log\left(2^{r_1}-1\right)+\frac{1}{2}\log\beta \right\rceil$ given maximum allowed rate loss, $r_1$, and fixed ADC resolution, $b_{AD}$.
It is revealed that if $\beta$ increases by four times, $b_{DA}$ should increase about one bit for maintaining the same rate loss.
While for low SNR, the rate loss is proven ignorable.

The rest of this paper is organized as follows. System model is described in Section \uppercase\expandafter{\romannumeral2}. In Section \uppercase\expandafter{\romannumeral3}, we derive an asymptotic expression for the achievable rate and analyze the rate loss due to quantization. Section \uppercase\expandafter{\romannumeral4} presents simulation results. Conclusions are drawn in Section \uppercase\expandafter{\romannumeral5}.

\emph{Notations}: $\textbf{A}^T$, $\textbf{A}^*$ and $\textbf{A}^H$ represent the transpose, conjugate and conjugate transpose of $\textbf{A}$, respectively. $\textbf{A}_{i,j}$ represents the $(i, j)$th element of $\textbf{A}$. $\textrm{tr}\{\textbf{A}\}$ denotes the trace of $\textbf{A}$ and $\textrm{diag}(\textbf{A})$ keeps only the diagonal entries of $\textbf{A}$.
$\mathbb{E}\{\cdot\}$ is the expectation operator. $\longrightarrow$ denotes almost sure convergence. $\lceil\cdot\rceil$ is the ceiling function.
\vspace{-0.4cm}

\section{System Model}
\subsection{System Description}
\begin{figure}[tb]
\centering\includegraphics[width=0.5\textwidth,bb= 15 56 595 210]{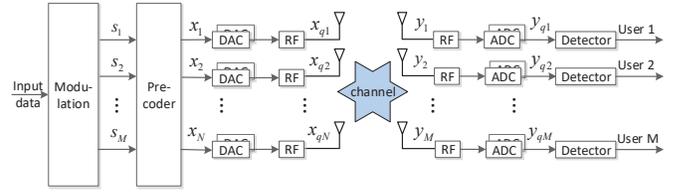}
\caption{Block diagram of multiuser massive MIMO downlink.}
\label{block}
\vspace{-0.5cm}
\end{figure}

Consider a downlink multiuser massive MIMO system illustrated in Fig. \ref{block}. The BS is equipped with $N$ antennas and serves $M$ single-antenna users simultaneously with user load ratio, $\beta \triangleq \frac{M}{N}<1$ in general.

At the transmitter, data vector $\textbf{s}\in \mathbb{C}^{M\times 1}$ with $\mathbb{E}\{\textbf{s}\textbf{s}^{H}\}=\textbf{I}_{M}$ is processed by $\textbf{P}\in \mathbb{C}^{N\times M}$. The precoded signal, i.e., $\textbf{x}\triangleq\textbf{Ps}$, is then passed through finite-bit DACs before transmission. The signal vector after DACs, $\textbf{x}_{q}\in \mathbb{C}^{N\times 1}$, is expressed as
\vspace{-0.2cm}
\begin{equation}
\label{xq}
\textbf{x}_{q}=\mathcal{Q}_{DA}(\textbf{x})=\mathcal{Q}_{DA}(\textbf{Ps}),
\end{equation}
where $\mathcal{Q}_{DA}(\cdot)$ denotes the quantization operation of DACs.

At the receiver, the received signal vector $\textbf{y}\in \mathbb{C}^{M\times 1}$ equals
\begin{equation}
\label{y}
\textbf{y}=\textbf{H}\textbf{x}_{q}+\textbf{n},
\vspace{-0.2cm}
\end{equation}
where $\textbf{n}\sim \mathcal{CN}(0,\sigma_n^2)$ is the additive white Gaussian noise (AWGN) and $\textbf{H}\in \mathbb{C}^{M\times N}$ is a Rayleigh fading channel whose entries are i.i.d. $\mathcal{CN}(0,1)$.
Let $\mathcal{Q}_{AD}(\cdot)$ denote the ADC quantization. The quantized received signal vector is
\begin{equation}
\label{yq}
\textbf{y}_{q}=\mathcal{Q}_{AD}(\textbf{y})=\mathcal{Q}_{AD}(\textbf{H}\textbf{x}_{q}+\textbf{n}).
\vspace{-0.1cm}
\end{equation}
%\vspace{-1.5cm}
\subsection{Quantization Models}
Since the quantization of both finite-bit DACs and ADCs are nonlinear, it is difficult to conduct exact characterizations on the quantization operations.
Fortunately, an approximate linear representation of quantization is available by the Bussgang theorem \cite{Bus1}.
For a Gaussian source data $\textbf{s}$, the quantization input signals can be considered as Gaussian as well \cite{Bus2}. Thus for the Gaussian input, we can apply the Bussgang theorem to decompose the quantized signal into two uncorrelated parts.
Taking ADC for example, we have
\begin{equation}
\label{Bus_AD}
\textbf{y}_{q}=\mathcal{Q}_{AD}(\textbf{y})=\textbf{F}\textbf{y}+\textbf{n}_{AD},
%\vspace{-0.2cm}
\end{equation}
where $\textbf{F}\in \mathbb{C}^{M\times M}$ is a diagonal matrix consisting of the quantization gains and $\textbf{n}_{AD}$ denotes the quantization noise uncorrelated with $\textbf{y}$.
Note that the diagonal entries of $\textbf{F}$ are considered equal here under the assumption of equal average power of each received signal which asymptotically holds because the rows of $\textbf{H}$ tend quasi-orthogonal with large $N$ in the massive MIMO setup.
According to [9, Eqs. (5) and (25)], \eqref{Bus_AD} can be reduced to
\begin{equation}
\label{ADC}
\textbf{y}_{q}=\mathcal{Q}_{AD}(\textbf{y})=(1-\rho_{AD})\textbf{y}+\textbf{n}_{AD},
\vspace{-0.2cm}
\end{equation}
where $\rho_{AD}=\frac{\mathbb{E}\{\|\textbf{y}_{q}-\textbf{y}\|^2\}}{\mathbb{E}\{\|\textbf{y}\|^2\}}$ denotes the distortion factor and [9, Eqs. (6), (25) and (28)]
\begin{equation}
\begin{aligned}
\label{cor_AD}
\mathbb{E}\{\textbf{n}_{AD}\textbf{n}_{AD}^H\}&=\rho_{AD}(1-\rho_{AD})\mathbb{E}\{\textrm{diag}(\textbf{y}\textbf{y}^H)\}.
\end{aligned}
\end{equation}
Let $b_{AD}$ be the quantization bits of ADC.
As in most existing works, e.g., \cite{ADC2}, \cite{Bus2} and \cite{ADC3}, we consider the optimal non-uniform ADCs since it provides a tractable and effective way of well characterizing the quantization performance.
For uniform quantizers used in practice, this can serve as a performance bound and the gap between the optimal and uniform ones is in general marginal, especially for popular quantization levels \cite{rho}.
The value of $\rho_{AD}$ is exemplified as \{0.3634, 0.1175, 0.03454, 0.009497, 0.002499, 0.0006642, 0.0001660, 0.00004151\} with $b_{DA}=\{1,2,3,4,5,6,7,8\}$ \cite{rho}.
In the condition of moderate to high-resolution quantizations, e.g., $b_{AD}\geq3$, we have \cite{Bus2}
\begin{equation}
\label{rho}
\rho_{AD}\approx\frac{\pi\sqrt{3}}{2}2^{-2b_{AD}}.
\end{equation}
While for DAC, the Bussang Theorem can also be applied  for making the operation linearly approximated as exemplified in \cite{DAC2} and \cite{DAC3}. Similarly, we have the following representation:
\begin{equation}
\label{DAC}
\textbf{x}_{q}=\mathcal{Q}_{DA}(\textbf{x})=\sqrt{1-\rho_{DA}}\textbf{x}+\textbf{n}_{DA},
\vspace{-0.1cm}
\end{equation}
where $\rho_{DA}$ is determined by $b_{DA}$ and
\begin{equation}
\begin{aligned}
\label{cor_DA}
\mathbb{E}\{\textbf{n}_{DA}\textbf{n}_{DA}^H\}=\rho_{DA}\mathbb{E}\{\textrm{diag}(\textbf{x}\textbf{x}^H)\}.
\end{aligned}
\end{equation}
Compared to ADCs, a scalar factor, $\frac{1}{\sqrt{1-\rho_{DA}}}$, is multiplied for DACs in order to guarantee that $\mathbb{E}\{\|\textbf{x}_q\|^2\}=\mathbb{E}\{\|\textbf{x}\|^2\}$.

\iffalse
\begin{table*}[tbh]
\centering
\caption{Parameter values for quantization model \cite{Bus2}}
\label{Parameters}
\begin{IEEEeqnarraybox}[\IEEEeqnarraystrutmode\IEEEeqnarraystrutsizeadd{2pt}{1pt}]{v/c/v/c/v/c/v/c/v/c/v/c/v/c/v/c/v/c/v}
\IEEEeqnarrayrulerow\\\IEEEeqnarrayseprow[2pt]\\
& \mbox{Resolution $b$ (bit)} && \mbox{1}&& \mbox{2}&& \mbox{3}&& \mbox{4}&& \mbox{5}&& \mbox{6}&& \mbox{7}&&\mbox{8}& \IEEEeqnarraystrutsize{0pt}{0pt}\\
\IEEEeqnarrayseprow[2pt]\\
\IEEEeqnarrayrulerow\\
\IEEEeqnarrayseprow[2pt]\\
& \mbox{Distortion factor $\rho$} && \mbox{0.3634} && \mbox{0.1175}&& \mbox{0.03454}&& \mbox{0.009497}&& \mbox{0.002499}&& \mbox{0.0006642}&& \mbox{0.0001660}&& \mbox{0.00004151}&
\IEEEeqnarraystrutsize{0pt}{0pt}\\
\IEEEeqnarrayseprow[2pt]\\
\IEEEeqnarrayrulerow
\end{IEEEeqnarraybox}
\end{table*}
\fi

\section{Asymptotic Performance Analysis}
%In the above we have described the approximate linear model of DAC and ADC quantization at the transceiver.
This section derives the asymptotic downlink achievable rate and characterizes the rate loss due to finite-bit DAC/ADCs under the assumption of large $N$ but fixed user load ratio $\beta$.

\subsection{Asymptotic User Rate}
Substituting \eqref{y} and \eqref{DAC} into \eqref{ADC}, the quantized signal received by users becomes
\begin{equation}
\begin{aligned}
\label{yq2}
\textbf{y}_{q}=&(1-\rho_{AD})\sqrt{1-\rho_{DA}}\textbf{H}\textbf{P}\textbf{s}+(1-\rho_{AD})\textbf{H}\textbf{n}_{DA}+\\
&\textbf{n}_{AD}+(1-\rho_{AD})\textbf{n}.
\end{aligned}
\end{equation}
Let $\textbf{h}_{k}^{T}$ denote the $k$th row of $\textbf{H}$. The received signal of the $k$th user equals
\begin{equation}
\begin{aligned}
\label{yk}
y_{k}=&(1-\rho_{AD})\sqrt{1-\rho_{DA}}\textbf{h}_{k}^{T}\textbf{P}\textbf{s}+(1-\rho_{AD})\textbf{h}_{k}^{T}\textbf{n}_{DA}+\\
&n_{AD,k}+(1-\rho_{AD})n_{k},
\end{aligned}
\end{equation}
where $n_{AD,k}$ and $n_{k}$ are, respectively, the $k$th element of $\textbf{n}_{AD}$ and $\textbf{n}$.
Then, the signal-to-interference, quantization and noise ratio (SIQNR) of the $k$th user, $\gamma_k$, can be expressed as
\begin{equation}
\begin{aligned}
\label{SIQNR1}
\gamma_k=\frac{S_k}{I_k+Q_{1k}+Q_{2k}+N_k},
\end{aligned}
\end{equation}
where $S_k, I_k, Q_{1k}, Q_{2k}$, and $N_k$ respectively denote the power of received signal, multiuser interference, DAC quantization noise, ADC quantization noise, and Gaussian noise. Obviously from \eqref{yk}, we have $N_k=(1-\rho_{AD})^2\sigma_n^2$.
Considering $\textbf{P}$ as a typical zero-forcing (ZF) precoder, asymptotic expressions of $S_k, I_k, Q_{1k}$ and $Q_{2k}$ under the assumption of large $N$ but fixed $\beta$ are, respectively, given as
\begin{equation}
\label{S}
S_k\longrightarrow(1-\rho_{AD})^2(1-\rho_{DA})P\left(\frac{1}{\beta}-1\right),
\end{equation}
\begin{equation}
\label{I}
I_k\longrightarrow0,
\end{equation}
\begin{equation}
\label{Q1}
Q_{1k}\longrightarrow(1-\rho_{AD})^2\rho_{DA}P,
\end{equation}
\begin{equation}
\begin{aligned}
\label{Q2}
Q_{2k} & \longrightarrow \rho_{AD}(1-\rho_{AD})(1-\rho_{DA})P\left(\frac{1}{\beta}-1\right)+\rho_{AD}\rho_{DA}\\
& \times (1-\rho_{AD})P+\rho_{AD}(1-\rho_{AD})\sigma_n^2,
\end{aligned}
\end{equation}
where the derivations are detailed in Appendix.
%Let $\bar{\gamma}(b_{DA},b_{AD})$ denote the asymptotic SIQNR with $b_{DA}$-bit DACs and $b_{AD}$-bit ADCs.
By substituting \eqref{S}-\eqref{Q2} into \eqref{SIQNR1}, the asymptotic SIQNR is given by
\begin{equation}
\begin{aligned}
\label{SIQNR3}
\bar{\gamma}(b_{DA},b_{AD})=
\frac{(1-\rho_{AD})(1-\rho_{DA})(\frac{1}{\beta}-1)\gamma_0}{\rho_{DA}\gamma_0+\rho_{AD}(1-\rho_{DA})(\frac{1}{\beta}-1)\gamma_0+1},
\end{aligned}
\end{equation}
where $\gamma_0=\frac{P}{\sigma_n^2}$ represents the average system SNR.
Supposed that the quantization and interference noise suffers the worst case, i.e., Gaussian distribution, the asymptotic achievable rate can be bounded by
\begin{equation}
\begin{aligned}
\label{rate2}
&R(b_{DA},b_{AD})=\log(1+\bar{\gamma}(b_{DA},b_{AD}))\\
&\!=\!\log\left(1\!+\!\frac{(1-\rho_{AD})(1-\rho_{DA})(\frac{1}{\beta}-1)\gamma_0}{\rho_{DA}\gamma_0\!+\!\rho_{AD}\!(\!1\!-\!\rho_{DA}\!)\!(\frac{1}{\beta}\!-\!1)\gamma_0\!+\!1}\right).
\end{aligned}
\end{equation}

\subsection{Quantization Rate Loss Analysis}
On the basis of \eqref{rate2}, we are ready to analyze the rate loss caused by both finite-bit DACs and ADCs. Considering three special cases, i.e., using ideal (infinite-bit) DACs or ideal ADCs, or both, as benchmarks, we have
\begin{equation}
\begin{aligned}
\label{rate_full_DAC}
R(\infty,b_{AD})=\log\left(1+\alpha_{AD}\overline{\gamma}_{ideal}\right),
\end{aligned}
\end{equation}
\begin{equation}
\begin{aligned}
\label{rate_full_ADC}
R(b_{DA},\infty)=\log\left(1+\alpha_{DA}\overline{\gamma}_{ideal}\right),
\end{aligned}
\end{equation}
\begin{equation}
\begin{aligned}
\label{rate_full_both}
R(\infty,\infty)=
\log\left(1+\overline{\gamma}_{ideal}\right).
\end{aligned}
\end{equation}
where $\overline{\gamma}_{ideal} \triangleq (\frac{1}{\beta}-1)\gamma_0$ is regarded as the nominal SNR with ideal converters at both sides. Here, $\alpha_{AD} \triangleq \frac{(1-\rho_{AD})}{\rho_{AD}(\frac{1}{\beta}-1)\gamma_0+1}$ and $\alpha_{DA} \triangleq \frac{(1-\rho_{DA})}{\rho_{DA}\gamma_0+1}$ are multipliers indicating the equivalent SNR degradation factors due to the use of finite-bit ADCs and DACs, respectively.

\textbf{Remark 1.}
\emph{Assuming the same resolution for ADCs and DACs, i.e. $b_{AD}=b_{DA}$, we observe that $\alpha_{DA}>\alpha_{AD}$ for typical user load values, $\beta<\frac{1}{2}$ in massive MIMO, which implies that ADC quantization always causes more pronounced SIQNR degradation than DAC in this case.}

Now by subtracting \eqref{rate2} from \eqref{rate_full_both}, the rate loss is directly characterized for analyzing the joint effect of both finite-bit DACs and ADCs on the achievable rate. However, the expression obtained appears too complicated to extract helpful insight.
In order to make the analysis more tractable, we alternatively first focus on a special case when $b_{AD}$ is fixed and study the behavior of varying $b_{DA}$ at BS on the rate loss.
By subtracting \eqref{rate2} from \eqref{rate_full_DAC}, the rate loss caused by finite-bit DACs can be characterized as
\begin{equation}
\begin{aligned}
\label{delta_rate_DAC}
&\Delta R_{DA}(b_{DA})=
\log\left(1+\frac{(1-\rho_{AD})(\frac{1}{\beta}-1)\gamma_0}{\rho_{AD}(\frac{1}{\beta}-1)\gamma_0+1}\right)\\
&-\log\left(1\!+\!\frac{(1-\rho_{AD})(1-\rho_{DA})(\frac{1}{\beta}-1)\gamma_0}{\rho_{DA}\gamma_0\!+\!\rho_{AD}\!(\!1\!-\!\rho_{DA}\!)\!(\frac{1}{\beta}\!-\!1)\gamma_0\!+\!1}\right),\\
\end{aligned}
\end{equation}
where $\rho_{DA}$ is a variable while $\rho_{AD}$ is a constant.
\iffalse
\begin{equation}
\begin{aligned}
\label{delta_rate_DAC}
\Delta R_{DA}(b_{DA},b_{AD})=\log\left(\frac{1+\alpha_{AD}\overline{\gamma}_{ideal}}{1+\overline{\gamma}}\right).
\end{aligned}
\end{equation}
\fi

For low SNRs with $\gamma_0\rightarrow0$, we have
\begin{equation}
\begin{aligned}
\label{delta_rate_DAC_low}
\Delta R_{DA}^{low}(b_{DA})\!=\!
\log\!\left(\!\frac{1+(1-\rho_{AD})(\frac{1}{\beta}-1)\gamma_0}{1\!+\!(1\!-\!\rho_{AD})(1\!-\!\rho_{DA})(\frac{1}{\beta}\!-\!1)\gamma_0}\!\right) \!\rightarrow\! 0,
\end{aligned}
\end{equation}
which implies that the rate loss due to DAC quantization can be ignorable at low SNRs even when low-resolution converters are employed at both sides.
Furthermore, we normalize $\Delta R_{DA}^{low}$ by $\gamma_0$ as
\begin{equation}
\begin{aligned}
\label{delta_rate_DAC_low2}
\frac{\Delta R_{DA}^{low}(b_{DA})}{\gamma_0}
\rightarrow \frac{1}{\ln2}\rho_{DA}(1-\rho_{AD})\left(\frac{1}{\beta}-1\right),
\end{aligned}
\end{equation}
where we use $(1+x)^{\frac{1}{x}}\rightarrow \textrm{e}$ when $x\rightarrow 0$.
It implies that the DAC quantization loss per energy is directly proportional to $\rho_{DA}$ and decreases with $\beta$ increasing.
Similar observations can be made with fixed $b_{DA}$ but varying $b_{AD}$.

For high SNRs with $\gamma_0\gg1$, we have
\begin{equation}
\begin{aligned}
\label{delta_rate_DAC_high}
&\Delta R_{DA}^{high}(b_{DA})=
\log \left(1+\frac{\frac{1}{\rho_{AD}}-1}{(\frac{1}{\rho_{DA}}-1)(\frac{1}{\beta}-1)+1}\right).
\end{aligned}
\end{equation}
It is obvious that the rate loss is in general not ignorable.
Given fixed $b_{AD}$, we therefore aim to determine the DAC resolution, $b_{DA}$, in order to guarantee the rate loss no larger than $r_1$ w.r.t the case using ideal DACs. Accordingly, characterizing $\Delta R_{DA}^{high}\leq r_1$ with \eqref{delta_rate_DAC_high} yields:
\begin{equation}
\begin{aligned}
\label{b_DAC}
b_{DA}&=\left\lceil\log \left[\frac{2}{\sqrt{3}\pi}\frac{(2^{r_1}-1)(\frac{1}{\beta}-1)}{(2^{r_1}\!-\!1)(\frac{1}{\beta}\!-\!1)\!+\!\frac{1}{\rho_{AD}}\!-\!2^{r_1}}\right]^{-\frac{1}{2}}\right\rceil\\
&\overset{(a)}\approx\left\lceil b_{AD}-\frac{1}{2}\log\left(2^{r_1}-1\right)-\frac{1}{2}\log\left(\frac{1}{\beta}-1\right) \right\rceil\\
&\overset{(b)}\approx\left\lceil b_{AD}-\frac{1}{2}\log\left(2^{r_1}-1\right)+\frac{1}{2}\log\beta \right\rceil,
\end{aligned}
\end{equation}
where $(a)$ utilizes \eqref{rho} with $\rho_{AD}\ll1$ for $b_{AD}\geq3$ and $(b)$ is under the fact that $\beta\ll1$ usually holds in massive MIMO.

\textbf{Remark 2.}
\emph{
It can be observed that $b_{DA}$ should increase about one bit if $r_1$ becomes two bit smaller. This is intuitively reasonable because the real and imaginary components are quantized separately.
Besides, $b_{DA}$ should increase with increasing $\beta$, i.e., serving more users or equipping less transmitting antennas. Higher DAC resolution retains more transmitting symbol information and consequently can serve more users.
%$b_{DA}$ is affected by user load ratio $\beta$ logarithmically.
In particular from \eqref{b_DAC}, $b_{DA}$ should increase about one bit if $\beta$ grows by four times.
For a special choice of $r_1=1$, $b_{DA}\approx\left\lceil b_{AD}+\frac{1}{2}\log\beta \right\rceil$.
}

On the other hand, given fixed-bit DACs, the resolution of ADCs can be analogously derived as
\begin{equation}
\begin{aligned}
\label{b_ADC}
b_{AD}\approx\left\lceil b_{DA}-\frac{1}{2}\log\left(2^{r_2}-1\right)-\frac{1}{2}\log\beta \right\rceil,
\end{aligned}
\end{equation}
where $r_2$ is the maximum allowed rate loss due to finite-bit ADCs.

\textbf{Remark 3.}
\emph{
Comparing \eqref{b_ADC} with \eqref{b_DAC}, the difference of characterizing $b_{AD}$ and $b_{DA}$ only lies on the sign of last item, $\frac{1}{2}\log\beta$.
When more transmitting antennas are employed at BS, higher ADC resolution is needed to detect more information with $\beta$ decreasing.
%Higher ADC resolution detects more receiving information and consequently matches more transmitting antennas at BS.
}

\iffalse
On the other hand, the rate loss caused by finite-bit ADCs can be characterized by subtracting \eqref{rate2} from \eqref{rate_full_ADC} as
\begin{equation}
\begin{aligned}
\label{delta_rate_ADC}
&\Delta R_{AD}(b_{DA},b_{AD})\!=\!\log\left(1\!+\!\frac{\rho_{AD}(1\!-\!\rho_{DA})^2(\frac{N}{M}\!-\!1)\gamma_0}{\rho_{DA}(1-\rho_{DA})\gamma_0+1}\right).\\
\end{aligned}
\end{equation}
When $\gamma_0$ is low or high, we have
\begin{equation}
\begin{aligned}
\label{delta_rate_ADC_low}
&\Delta R_{AD}^{low}(b_{DA},b_{AD})=
\log\left(1\!+\!\rho_{AD}(1\!-\!\rho_{DA})^2(\frac{N}{M}\!-\!1)\gamma_0\right),
\end{aligned}
\end{equation}
and
\begin{equation}
\begin{aligned}
\label{delta_rate_ADC_high}
&\Delta R_{AD}^{high}(b_{DA},b_{AD})=
\log \left(1\!+\!\rho_{AD}(\frac{1}{\rho_{DA}}\!-\!1)(\frac{N}{M}\!-\!1)\right).
\end{aligned}
\end{equation}
When $\gamma_0$ is small, $\Delta R_{AD}^{low}\approx0$ and the ADC quantization noise is ignorable.

Similarly, given fixed-bit DACs and the maximum rate loss allowed due to finite-bit ADCs, i.e., $r_2$, the resolution of ADCs is
\begin{equation}
\begin{aligned}
\label{b_ADC}
b_{AD}%&=\left\lceil \log \left[\frac{2}{\sqrt{3}\pi}\frac{(2^{\Delta R_{AD}^{high}}-1)}{(\frac{1}{\rho_{DA}}-1)(\frac{N}{M}-1)}\right]^{-\frac{1}{2}}\right\rceil,\\
&\approx\left\lceil b_{DA}-\frac{1}{2}\log\left(2^{r_2}-1\right)+\frac{1}{2}\log\left(\frac{1}{\beta}-1\right) \right\rceil.
\end{aligned}
\end{equation}
\fi

\section{Simulation Results}
\begin{figure}[tb]
\centering\includegraphics[width=0.36\textwidth,bb=50 230 550 580]{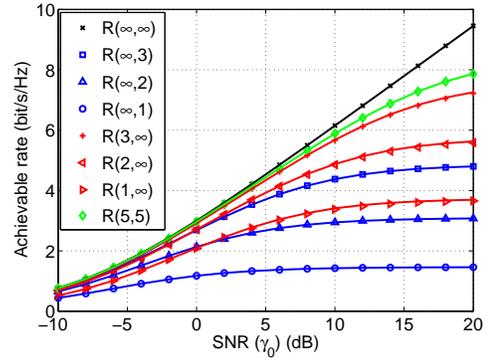}
\caption{Achievable rate with finite-bit DACs and ADCs ($\beta=\frac{1}{8}$).}
\label{comparison}
\vspace{-0.4cm}
\end{figure}

In this section, we test the rate under the assumption of large $N$ but fixed $\beta$.
Fig.~\ref{comparison} compares the achievable rate with infinite and finite-bit DAC/ADCs. Markers correspond to simulation results while solid lines correspond to the derived expressions.
We observe that finite-bit ADCs cause more rate loss than DACs with the same resolution, which verifies \emph{Remark~1}.
Rate loss at low SNRs is rather marginal because thermal noise is dominating in this case.
%Besides, the rate loss with 5-bit DACs and ADCs is rather small compared to the ideal case. It means that low resolution converters can achieve comparable performance to ideal ones.

\begin{figure}[tb]
\centering\includegraphics[width=0.36\textwidth,bb=50 230 550 580]{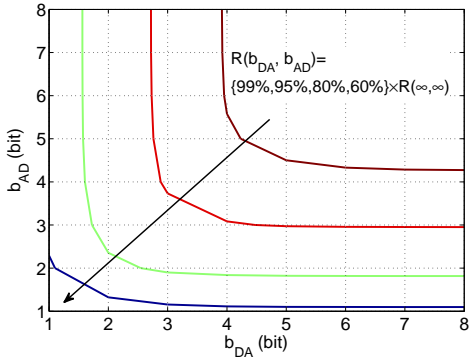}
\caption{Contour of asymptotic rate with finite-bit DACs and ADCs ($\beta=\frac{1}{8},~\gamma_0=-10$dB).}
\label{contour}
\end{figure}

%Specifically, $R(2,1)>R(1,2)$. This means that one more bit for DAC contributes more to achievable rate than ADC with low $\gamma_0$ because the operation of DAC is irrelevant to AWGN.

\begin{figure}[tb]
\centering\includegraphics[width=0.36\textwidth,bb=50 230 550 550]{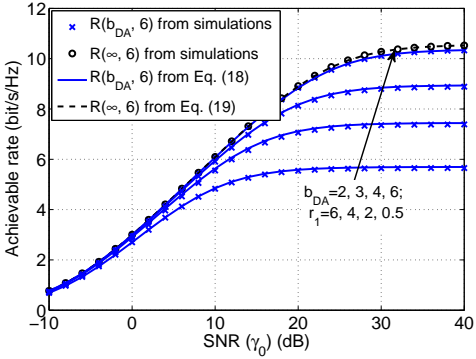}
\caption{Rate loss by finite-bit DACs given fixed-bit ADCs and $\beta=\frac{1}{8}$.}
\label{full_DAC}
\vspace{-0.4cm}
\end{figure}
Fig. \ref{contour} displays the contour of asymptotic rate. Obviously, $R(b_{DA},b_{AD})$ increases when both $b_{DA}$ and $b_{AD}$ grow. We find that 99\%$R(\infty,\infty)$ can be achieved by using 5-bit DACs and ADCs because AWGN is rather more pronounced than quantization noise when $\gamma_0=-10$dB. In other words, low resolution converters are suitable to low SNR conditions as expected.
Fig. \ref{full_DAC} shows the achievable rate with varying $b_{DA}$ and fixed $b_{AD}$. Given $\beta=\frac{1}{8}$, $b_{AD}=6$ and $r_1=6, 4, 2, 0.5$, we have $b_{DA}=2, 3, 4, 6$ according to \eqref{b_DAC}. We observe that the rate loss caused by finite-bit DACs  at high SNR decreases about two bits with $b_{DA}$ increasing one bit when $b_{DA}$ is small.

\section{Conclusion}
We derive the asymptotic downlink rate affected by both DAC and ADC quantizations in multiuser massive MIMO and reveal that the resolution of DACs should increase with the user load ratio in order to maintain a desired rate loss.
%At low SNR, low resolution DAC/ADCs achieve comparable performance to ideal ones.

\appendix
From \eqref{yk}, the evaluation of $S_k$ and $I_k$ are, respectively,
\begin{equation}
\begin{aligned}
\label{app_S}
S_k=(1-\rho_{AD})^2(1-\rho_{DA})|\textbf{h}_{k}^{T}\textbf{p}_k|^2,
\end{aligned}
\end{equation}
\begin{equation}
\begin{aligned}
\label{app_I}
&I_k=(1-\rho_{AD})^2(1-\rho_{DA})\sum_{j\neq k}|\textbf{h}_{k}^{T}\textbf{p}_j|^2.
\end{aligned}
\end{equation}
Then using \eqref{yk} and substituting \eqref{cor_DA}, we have
\begin{equation}
\begin{aligned}
\label{app_Q1}
Q_{1k}&=(1-\rho_{AD})^2\textbf{h}_{k}^{T}\mathbb{E}\{\textbf{n}_{DA}\textbf{n}_{DA}^H\}\textbf{h}_{k}^{*}\\
&=(1-\rho_{AD})^2\rho_{DA}\textbf{h}_{k}^{T}\textrm{diag}(\textbf{P}\textbf{P}^{H})\textbf{h}_{k}^{*},
\end{aligned}
\end{equation}
where we use $\textbf{x}=\textbf{Ps}$ and $\mathbb{E}\{\textbf{s}\textbf{s}^{H}\}=\textbf{I}_{M}$.
Similarly,
\begin{equation}
\begin{aligned}
\label{app_Q2}
&Q_{2k}=\mathbb{E}\{|n_{AD,k}|^2\}\overset{(a)}=\rho_{AD}(1-\rho_{AD})\mathbb{E}\{\textrm{diag}(\textbf{y}\textbf{y}^{H})_{k,k}\}~~~~~~~~~~~~~~~~~~~~~~~\\
&\overset{(b)}=\rho_{AD}(1-\rho_{AD})\mathbb{E}\{\textrm{diag}((\textbf{H}\textbf{x}_{q}+\textbf{n})(\textbf{H}\textbf{x}_{q}+\textbf{n})^H)_{k,k}\}\\
&\overset{\!(c)\!}=\!\rho_{AD}\!(\!1\!-\!\rho_{AD}\!)
\mathbb{E}\!\{\!\textbf{h}_{k}^{T}[\!(\!1\!-\!\rho_{DA}\!)\textbf{x}\textbf{x}^{H}\!\!+\!\rho_{DA}\textrm{diag}(\textbf{x}\textbf{x}^{H}\!)\!]\textbf{h}_{k}^{*}\!+\!\!|\!n_k\!|^2\}\\
&\overset{(d)}=\rho_{AD}(1\!-\!\rho_{AD})(1\!-\!\rho_{DA})\sum_{j=1}^M|\textbf{h}_{k}^{T}\textbf{p}_j|^2\!+\!\rho_{AD}(1\!-\!\rho_{AD})\sigma_n^2\\
&~~~+\rho_{AD}(1-\rho_{AD})\rho_{DA}\textbf{h}_{k}^{T}\textrm{diag}(\textbf{P}\textbf{P}^{H})\textbf{h}_{k}^{*},\\
\end{aligned}
\end{equation}
where $(a)$, $(b)$ and $(c)$ use \eqref{cor_AD}, \eqref{y} and \eqref{DAC}, respectively, and $(d)$ follows from $\textbf{x}=\textbf{Ps}$ and $\mathbb{E}\{\textbf{s}\textbf{s}^{H}\}=\textbf{I}_{M}$.

Now consider $\textbf{P}$ designed by the popular ZF precoder, i.e., $\textbf{P}=\sqrt{\frac{P}{\textrm{tr}\{(\textbf{H}\textbf{H}^H)^{-1}\}}}\textbf{H}^H(\textbf{H}\textbf{H}^H)^{-1}$,
%\begin{equation}
%\label{zf}
%\textbf{P}=\sqrt{\frac{P}{\textrm{tr}\{(\textbf{H}\textbf{H}^H)^{-1}\}}}\textbf{H}^H(\textbf{H}\textbf{H}^H)^{-1},
%\end{equation}
where $P$ is the total transmit power.
Given that $\textbf{H}\textbf{H}^H$ is a complex Wishart matrix, we have $\textrm{tr}\{(\textbf{H}\textbf{H}^H)^{-1}\}\longrightarrow \frac{M}{N-M}$ under the assumption of large $N$ but fixed $\beta$ \cite{Wishart}. Combining the definition of $\textbf{P}$, we have
\begin{equation}
\label{wishart3}
\textbf{H}\textbf{P}\longrightarrow \sqrt{P\left(\frac{1}{\beta}-1\right)}\textbf{I}_M.
\end{equation}
Then, the desired asymptotic expressions \eqref{S} and \eqref{I} follow from \eqref{app_S} and \eqref{app_I} by substituting \eqref{wishart3}.
As for $Q_{1k}$ and $Q_{2k}$, the diagonal entries of $\textrm{diag}(\textbf{P}\textbf{P}^{H})$ tend to the same with large $N$. Since $\textrm{tr}\{\textbf{P}\textbf{P}^{H}\}\!=\!P$, we have
$\textrm{diag}(\textbf{P}\textbf{P}^{H})\!\longrightarrow\!\frac{P}{N}\textbf{I}_N$. Finally, the desired expressions \eqref{Q1} and \eqref{Q2} follow from \eqref{app_Q1} and \eqref{app_Q2} by using $\frac{1}{N}\textbf{h}_{k}^{T}\textbf{h}_{k}^{*}\!\longrightarrow \!1$ due to the Central Limit Theorem.

\end{document}